\documentstyle[11pt]{article}
\begin{document}
\title{Dilatonic formulation for conducting cosmic \\string models}
\author{ {\bf Brandon Carter}\\ 
D.A.R.C. (CNRS), Observatoire de Paris, 92 Meudon, France}
\date{Contrib. to JR99 meeting, Weimar, September 1999,\\ 
to be published in Annalen der Physik (Leipzig).}

\maketitle
{\bf Abstract.\ }
  It is shown how the the introduction of a suitably defined dilatonic
auxiliary field, $\Phi$ say, makes it possible for the non-linear Lagrangian 
for a generic elastic string model, of the kind appropriate for representing 
superconducting cosmic strings, to be converted into a standardised form as 
the sum of a kinetic term that is just homogeneously quadratic in the relevant
scalar phase gradient (as in a simple linear model) together with a potential
energy term, $V$ say, that is specified as a generically non-linear function
of $\Phi$. The explicit form of this function is derived for various 
noteworthy examples, of which the most memorable is that of the transonic 
string model, as characterised by a given mass scale, $m$ say, for which
this potential energy density will be expressible in terms of the zero
current limit value $\Phi_{\!_0}$ of $\Phi$ by $ V= {1\over 2}\,  m\,  
\big(\Phi_{\!_0}^{-2} \Phi^2+  \Phi_{\!_0}^{\,2}\, \Phi^{-2}  \big)$

\vskip 1 cm

{\bf Introduction}
\label{intro}

The purpose of this article is to show how the introduction of an
appropriate dilatonic auxiliary field, $\Phi$ say can be used to
provide an elegant reformulation of the action variation principle
governing the dynamics of a generic member of the category of
``superconducting'' cosmic string models of the simple elastic
kind~\cite{C89a,C89b} needed for the macroscopic
description~\cite{CP95} of the effect~\cite{Peter92a} of the Witten
mechanism~\cite{Witten85} whereby some kind of conserved current is
localised on vacuum vortex defects. (Although there is no reason why
several independent currents should not occur simultaneously
~\cite{C94b}, this discussion will be concerned only with cases in
which only a single dominant current is taken into account, which can
be expected to be a good approximation for a wide range of
applications.)

This work has been motivated by the consideration that the variation
principles governing the dynamics of typical theoretical physical
models are commonly characterised by the desirable property of having
kinetic terms that are just quadratic. Explicitly this means that the
action has only (homogeneous or inhomogeneous) quadratic dependence on
the relevant field gradients, a feature that is highly desirable for
many technical purposes (particularly quantisation procedures). An
objective of the present work is to overcome the drawback that such a
convenient quadraticity feature was lacking in the kind of worldsheet
action principle originally developed by the present author~\cite{C89a}
for the purpose of describing the effects~\cite{CP95} of Witten type
currents in cosmic strings, though it did characterise the more
primitive conducting string model originally proposed for this purpose
by Witten himself~\cite{Witten85}.

For the purpose of describing the effect of his bosonic 
``superconductivity'' mechanism, the trouble with the primitive 
Witten model was that it was too restricted to be quantitatively or 
even qualitatively realistic. Its most notable inadequacy was that it 
implied subsonic wiggle propagation, whereas detailed analysis by 
Peter~\cite{Peter92a} showed that the extrinsic wiggle type
perturbations could actually be expected to propagate supersonically, 
meaning faster than the sonic type (longitudinal) perturbations of the 
condensate. This essential limitation of the original Witten string 
model, in which sonic perturbations could only propagate at the speed 
of light, was overcome by the introduction of a general category of 
elastic string models~\cite{C89a,C89b} governed by actions in which 
the dependence on the gradient of the relevant scalar phase field 
variable $\varphi$ is of a generically non-quadratic kind: the
non-quadratic phase gradient dependence is specified by a 
corresponding equation of state, which can be chosen~\cite{CP95} in 
such a way as to provide a very satisfactory description of the 
behaviour predicted by detailed microscopic analysis~\cite{Peter92a} 
of the bosonic condensate. Another application of this category of 
non-quadratic string models is to the case~\cite{C90,C95} that 
represents the averaged effect of wiggles~\cite{Vilenkin90,Martin95} 
on an underlying string of non-conducting Nambu-Goto type, which is 
characterised by the mathematically convenient property of 
transonicity, meaning that all kinds of propagations travel at the 
same (generically subluminal) speed.

Although it provides a misleading description of the effect of his 
bosonic ``superconducting'' string current mechanism, Witten's 
primitive purely quadratic string model  is more plausible, and 
certainly  the best presently available candidate, for the purpose 
of describing the effect of an alternative string current 
mechanism~\cite{Witten85} that he also proposed, namely that of 
fermionic zero mode excitations (for which, since no Cooper type 
condensation mechanism is involved, the term ``superconductivity'' 
is definitely not appropriate) at least in the weak current limit. 
(Allowance for the corrections needed for the treatment of stronger 
currents would require a sophisticated analysis presumably based on 
second quantisation which has not yet been carried out.)

Despite the facts that its physical validity seems to be limited to 
the case of weak fermionic currents, and that it lacks the 
mathematically convenient features (including exact integrability in 
flat space) of the idealised transonic model~\cite{C95}, the original 
Witten model~\cite{Witten85} has retained remarkably widespread 
popularity among superconducting string theorists, partly, it would 
seem, because of its (deceptively) attractive feature of being 
governed by an action with purely quadratic gradient dependence. 

The purpose of the present work is to show that this attractive feature 
is not restricted to the primitive Witten model but can be extended to 
the generic non linear category, including the transonic case, by a 
reformulation of the action whereby the original phase scalar $\varphi$ 
is supplemented in an unambiguously natural way by the introduction of 
an auxiliary dilatonic field $\Phi$ say, which allows the nonlinearity 
to be transferred from the kinetic term to a new potential energy term 
that is given as an appropriately nonlinear function, $V$ say, of 
$\Phi$.

The procedure used to obtain this reformulated action is adapted
from an analogous treatment of perfect fluid mechanics~\cite{C94a} that
has recently been shown to be particularly useful for the macroscopic
description of a rotating relative superfluid~\cite{CL95c} such
as is predicted to occur in neutron stars.

The result of the reformulation is to obtain a string world sheet 
Lagrangian function ${\cal L}$ that is generically expressible, using 
a vertical bar to denote gauge covariant differentiation with respect 
to worldsheet coordinates $\sigma^i$ ($i=0,1$), in the standardised form
 \begin{equation}
 {\cal L}=-{_1\over^2}\Phi^2\gamma^{ab}\varphi_{|a}
 \varphi_{|b}-V \label{0}\end{equation}
where $\gamma^{ab}$ is the inverse of the induced metric on the 
worldsheet. The latter will be given, in terms of the spacetime metric 
$g_{\mu\nu}$ with respect to four (or possibly higher) dimensional 
background coordinates $x^\mu$ by 
\begin{equation}
\gamma_{ab}=
g_{\mu\nu}x^\mu_{\, ,a}x^\nu_{\, ,b}\, , \label{gamm}\end{equation}
 using a comma to denote simple partial differentiation. 
If the current is characterised by an electromagnetic charge 
coupling constant $e$ then in the presence of an electromagnetic 
background field with gauge covector $A_\mu$ the corresponding 
gauge covariant derivative of the phase field $\varphi$ 
will be given by 
\begin{equation}\varphi_{|a}= \varphi_{,a}-eA_\mu x^\mu_{\, ,a}\, . 
\label{gradf}\end{equation}
Within the category of models characterised by a Lagrangian of the
form (\ref{0}), the primitive Witten model is specifiable as the
2-parameter sub-category that is obtained by application of a 
restricted action principle according to which the phase variable 
$\varphi$ is considered to be freely variable, but the dilatonic 
scalar $\Phi$ is treated as a constant, as also therefore is the 
potential $V$ (whose functional form has no significance in 
this case). Within this Witten subcategory, the degenerate limit of 
the Nambu-Goto model is obtained by imposing the further restriction 
that the fixed value for $\Phi$ should be zero (so in that particular 
-- non conducting -- case the phase variable $\varphi$ will be redundant).

Unlike the special Witten case, the generic case~\cite{C89a,C89b}
can be seen to be obtainable from (\ref{0}) by application of an 
unrestricted action principle in which both $\varphi$ and $\Phi$ (as 
well of course as the supporting world sheet) are considered to be 
freely variable, which evidently leads to dynamical equations whose 
particular form depends on the specific choice of the potential 
energy density scalar $V$ as a function of $\Phi$.

As an important semi-degenerate special case, this general category 
of elastic string models includes the one for which the 
potential energy function $V$ is simply taken to be 
constant: this special case is that of the chiral string 
model~\cite{CP99}, in which the dynamic equation obtained from 
the variation of $\Phi$ reduces to the condition that the gauge 
covariant phase gradient should be everywhere null, i.e. 
$\gamma^{ab}\varphi_{|a} \varphi_{|b}=0$. In contrast with the 
generic case for which the phase scalar $\varphi$ has two degrees of 
freedom at each point, in this chiral model there is only a single 
degree of phase field freedom at each point, corresponding (depending 
on the parity convention adopted for the string worldsheet 
orientation) to exclusively ``left'' (or ``right'') directed 
longitudinal propagation. Such a chiral string model is of physical 
relevance for application to a case of potential cosmological 
importance~\cite{C99,CD00} that arises naturally (particularly in the 
context of supersymmetry~\cite{DDT97,DDT98}) as a specialised variant 
of Witten's fermionic conductivity mechanism, namely the case in which 
the relevant zero modes are restricted to to propagate only along a 
single (e.g. purely ``left'') 
null propagation direction. Like the (non-degenerate) 
transonic~\cite{C95} string model (which will be presented explicily 
below) the (semi-degenerate) chiral string model shares with the well 
known case of the (fully degenerate, i.e. internally structureless) 
Nambu Goto string model the convenient property its equations of 
motion are explicitly integrable~\cite{CP99} in a flat space 
background.

\section
{The traditional formulation}
\label{trad}

Previous work on generic elastic string 
models~\cite{C89a,C89b,C95a,C97}
has been based on a variational principle specified in terms of an 
action integral of the form 
\begin{equation}
 {\cal I}=\int {\cal L}\, d{\cal S}\, ,
 \label{1} \end{equation} 
in which the element $d{\cal S}$ represents the induced surface measure 
on the world sheet, i.e.
\begin{equation}
 {\cal S}=\Vert \gamma\Vert^{1/2}\,d\sigma^{_0}\,d\sigma^{_1}  
 \label{2}\end{equation} 
where $\sigma^a$ ($a=0,1$) are worldsheet coordinates and 
$\vert\gamma\vert$ is the determinant of the induced metric (\ref{gamm}). 
The Lagrangian scalar ${\cal L}$ here is a function that, in the usual 
formulation,  depends only on the magnitude of the gauge covariant
worldsheet gradient (\ref{gradf}) of the scalar worldsheet potential 
$\varphi$, i.e. it is a function only of the scalar 
\begin{equation}
w=\kappa_{_0}\varphi_{\vert a}\varphi^{\vert a}\, ,
 \label{5} \end{equation}
in which $\kappa_{_0}$ is an adjustable constant that is used to obtain
a standard normalisation as described below.

The particular kind of model originally proposed by Witten himself is 
the subcategory for which the dependence of ${\cal  L}$ on the scalar
$\varphi_{\vert a}\varphi^{\vert a}$ is linear:
\begin{equation}
{\cal L}=-{_1\over^2}w- m^2\, ,
 \label{4} \end{equation}
where $m$ is a scale constant interpretable the Kibble mass 
characterising the isotropic Goto-Nambu type 
state of the string in the limit of vanishing current, i.e. the limit 
for which the scalar $\varphi$ is uniform.

Witten's simple linear model (\ref{4}) still seems the best available for 
the ordinary fermionic case, for which it makes sense as the obvious
first approximation in the weak current limit for which the fermions are 
sufficiently sparcely distributed for their effect on each other and on 
the string background to be negligible. Allowance for the corrections 
due to such effects would require a sophisticated analysis presumably 
based on second quantisation which has not yet been carried out.

The situation is different in the bosonic case, for which the background 
will be significantly affected by the presence of the condensate even in 
the zero current limit, a complication that is conveniently compensated 
by the possibility of dealing with it adequately on the basis just of a 
first quantised analysis. For the particular case of the simplified 
field theoretical model originally proposed by Witten, such an analysis 
has actually been carried out\cite{PBS88,Peter92a}, and has shown that 
for essential purposes, such as calculation of characteristic propagation 
speeds of small perturbations, the linear model (\ref{4}) is quite 
inadequate for describing the bosonic case. A satisfactory description 
can however be provided within the framework of the more general 
category of elastic string models, which were originally 
developed\cite{C89a,C89b} in terms of a variation principle for which 
the Lagrangian ${\cal L}$ in (\ref{1}) is given by an appropriate 
``equation of state'' as a non linear function of the variable $w$ 
defined by (\ref{2}).

Since only gradients, but not the absolute values, of $\varphi$
are involved, such a Lagrangian function will determine a corresponding 
conserved particle current vector, $z^a$ say, in the worldsheet, 
according to the Noetherian prescription
\begin{equation}
z^a=- {\partial {\cal L}\over\partial \varphi_{|a} }\ ,
\end{equation}
which implies
\begin{equation}
{\cal K}z^a= \kappa_{_0} \varphi^{|a}\ ,
\label{zcur}\end{equation}
(using the induced metric for internal index raising) where  ${\cal K}$ is
given as a function of $w$ by setting
\begin{equation}{1\over{\cal K}}=
-2{d{\cal L}\over dw} \ .
\label{calk}\end{equation}
This current $z^a$ {\it in} the worldheet can be represented by the
corresponding tangential current vector $z^\mu$  {\it on} the worldsheet,
where the latter is defined with respect to the background coordinates,
$x^\mu$, by $z^\mu=z^a x^\mu_{\, ,a}$.

The purpose of introducing the dimensionless scale constant $\kappa_{_0}$ is
to simplify macroscopic dynamical calculations by arranging for the variable
coefficient ${\cal K}$ to tend to unity when $w$ tends to zero, i.e. in the
limit for which the current is null. To obtain the desired simplification it
is convenient not to work directly with the fundamental current vector $z^\mu$
that (in units such that the speed of light and
the Dirac Planck constant $\hbar$ are set to unity)
will represent the quantized particle flux, but to work instead with a
corresponding rescaled particle current $c^\mu$ that is got by setting
\begin{equation}
z^\mu=\sqrt{\kappa_{_0}}\, c^\mu \ .
\label{scur}\end{equation}
In terms of the squared magnitude  $\chi=c^\mu c_\mu$ 
of this rescaled current $c^\mu$, the primary state variable $w$ defined 
by (\ref{5}) will be given simply by $w={\cal K}^2\chi$.
It is to be remarked that in the gauge coupled case, i.e. if $e$ is
non zero, there will be a corresponding electromagnetic current
vector obtained by a prescription of the usual form $j^\mu=\partial
{\cal L}/\partial A_\mu$ which simply gives $j^\mu=e z^\mu$
$=e\sqrt{\kappa_{_0}} c^\mu$.

The complete system of dynamical equations can conveniently be expressed
in terms of the surface stress momentum energy density tensor 
given~\cite{C95a,C97} by the formula
\begin{equation}
 \overline T{^{\mu\nu}}=2 {\partial {\cal L}\over \partial g_{\mu\nu}}
+{\cal L}\eta^{\mu\nu} \, , \label{genstress}
\end{equation}
using the notation 
\begin{equation}
\eta^{\mu\nu}= \gamma^{ab} x^\mu_{,a} x^\nu_{,b} 
 \end{equation}
for what is interpretable as the (first) fundamental tensor of the 
worldsheet. Independently of the particular form of the Lagrangian, the 
equations of motion obtained from the action (\ref{1}) will be expressible 
in the standard form~\cite{C95a,C97}
\begin{equation}
\overline\nabla_{\!\mu}\overline T{^\mu}{_\nu}=\overline f_\nu \ ,
\label{motion} \end{equation}
in which  $\overline f_\mu$ is the external force density acting on the 
worldsheet, and in which $\overline\nabla_{\!\mu}$ denotes the operator 
of surface projected covariant differentiation, as formally defined by
\begin{equation}
\overline\nabla{^\mu}\equiv\eta^{\mu\nu} \nabla_{\!\nu}
\equiv x^\mu_{\, ,a}\gamma^{ab}\nabla_{\! b}
\end{equation}
where $\nabla$ is the usual operator of covariant differentiation with 
respect to the Riemannian background connection. When the effect of 
electromagnetic coupling is significant the corresponding force density
$f_\mu$ will be given in terms of the field $F_{\mu\nu}=
A_{\nu,\mu}-A_{\mu,\nu}$ by $\overline f_\mu=eF_{\mu\nu} z^\nu$. 
The formula (\ref{genstress}) for the string surface stress tensor 
$\overline T{^{\mu\nu}}$ (from which the surface energy density $U$ and 
the string tension $T$ are obtainable as the negatives of its non-vanishing 
eigenvalues) can be seen to give a result having the simple form
 \begin{equation}
\overline T{^{\mu\nu}}={\cal L}\eta^{\mu\nu} +{\cal K} c^\mu c^\nu \, .
\label{stress}  \end{equation}
Even if the force density $f_\mu$ is 
non zero, its contraction with the current vector $z^\mu$, or with 
the corresponding rescaled current vector $c^\mu$, will vanish, and 
hence it can be seen from the preceeding formulae that the equations 
of motion (\ref{motion}) automatically imply the surface current 
conservation law
\begin{equation}
\overline\nabla_{\!\mu} c^\mu=0 \ . 
\end{equation}

The formulation presented above is the natural 
adaptation to strings of the Clebsch type variational formulation 
for relativistic fluid dynamics~\cite{Schutz70} in which the requisite 
Lagrangian scalar is interpretable as the pressure function $P$ say. 
It is well known that the Clebsch formulation is related via a 
generalised Legendre transformation to a corresponding Taub type 
variational formulation, in which it is the flow world lines 
that are treated as free variables, and in which instead of the 
pressure function $P$ the relevant Lagrangian is interpretable as the 
relativistic mass-energy density function $\rho$ say.
In an an analogous manner, in the elastic string case, the formulation
presented above in terms of ${\cal L}$ can be replaced by an
equivalent dually related formulation~\cite{C89a,C95a,C97} for which, 
instead of the phase potential $\varphi$, the independent variable is 
an approriately defined stream funcion $\tilde\varphi$ say, and for 
which the dual Lagrangian, $\Lambda$ say, is obtainable from the 
original Lagrangian function ${\cal L}$ by the Legendre type 
transformation formula
\begin{equation}
\Lambda={\cal L}+{w\over {\cal K}}.
\label{Lamb} \end{equation}
In the timelike current range where $w<0$ the tension and energy
density will be respectively given by  $T=-{\cal L}$, $U=-\Lambda$,
whereas in the spacelike range where $w>0$ they will be given by 
$T=-\Lambda$, $U=-{\cal L}$.  In either case the extrinsic
perturbation (``wiggle'') speed $c_{_{\rm E}}$ and the longitudinal
perturbation speed $c_{_{\rm L}}$ will be given~\cite{C89b,C97}
(relative to the preferred frame that exists in all except the 
``chiral'' case) by
\begin{equation} c_{_{\rm E}}^{\, 2}={T\over U}\, ,\hskip 1 cm
c_{_{\rm L}}^{\, 2}=-{dT\over dU}\, .\label{speeds}\end{equation}

\section{The dilatonic reformulation}
\label{reformulation}

The purpose of this article is to present a new formulation in terms of 
an alternative kind of action principle that has recently been developed 
in the context of ordinary relativistic fluid theory\cite{C94a} with a 
view to generalisation to the macroscopic treatment of 
superfluidity\cite{CL95c}. In this new treatment, the action depends not 
just on the gradient of the phase variable $\varphi$ but also on an 
auxiliary ``dilatonic'' amplitude variable $\Phi$. The introduction of 
this auxiliary variable allows the kinetic term in the Lagrangian to 
retain its traditional homogeneously quadratic dependence on the phase 
gradient, while the essential non-linearity of the model is encapsulated 
in a potential function $V$ that is specified as an appropriately
non-linear function of $\Phi$, in terms of which the total Lagrangian 
scalar takes the generic form (\ref{0}).

For an ordinary perfect fluid, the most important examples are the
homogeneously quadratic case, $V\propto \Phi^2$ which corresponds to the 
case of a ``dust'' type fluid for which the pressure $P$ is zero, and the
homogeneously quartic case, $V\propto \Phi^4$, which corresponds to the
``radiation gas'' case for which the pressure is related to the mass energy
density $\rho$ by the familiar relation $3P=\rho$. It is of course 
to be remarked that in the case of an ordinary perfect fluid a Lagrangian 
of the simple form (\ref{0}) can describe only irrotational motion, so 
further technical complications are needed for a fully generic
treatment\cite{C94a,CL95c}. However no such difficulty arises in the string
case with which we are concerned here because, as remarked above, there is
simply no room for rotation in a 1+1 dimensional worldsheet.

In the case of the elastic string models dealt with here, it can be seen 
by analogy with the fluid case~\cite{C94a,CL95c} that the required 
transformation to the standard form (\ref{0}) will be obtained by taking 
the dilatonic amplitude $\Phi$ to be given, as a function of $w$, by
\begin{equation}
 \Phi^2={\Phi_{\!_0}^2\over {\cal K}}\, ,
\label{dilam} \end{equation}
where its zero current value is given by the normalisation factor
\begin{equation}
 \Phi_{\!_0}^2=\kappa_{_0}\, , \end{equation}
while the required potential will have the (manifestly self dual)
form
\begin{equation}
 V=-{ {\cal L}+\Lambda\over 2}=w{d{\cal L}\over d w}-{\cal L}
\, ,\label{V}
 \end{equation}
whose derivative will be given by
 \begin{equation}
{dV\over d\Phi}= -{w\Phi\over \Phi_{\!_0}^2}\, .
\label{w} \end{equation}
It can thereby be seen from the form (\ref{0}) of the Lagrangian that the 
original defining relation (\ref{5}) for $w$ will be recovered as the 
condition for invariance of the action with respect to $\Phi$. Subject to 
this ``on shell'' identification, the surface stress energy momentum 
tensor $\overline T{^{\mu\nu}}$ obtained from the reformulated Lagrangian 
(\ref{0}) will be given by the same formula (\ref{stress}) as in the 
traditional version, so it follows that the ensuing dynamical equations 
will also be equivalent.

The reformulation can of course be inverted. Starting from the new 
formulation as characterised by a Lagrangian of the form (\ref{0}) for 
some given function $V$ of $\Phi$, the corresponding expression for 
${\cal L}$ as a function of $\Phi$, and hence implicitly, via (\ref{w}) 
of $w$, will be given by

\begin{equation}
{\cal L}={\Phi\over 2}{dV\over d\Phi}-V\, , \end{equation}
while the corresponding formula for the dual Lagrangian $\Lambda$ as a 
function of $\Phi$ (and hence implicitly via (\ref{dilam}) 
as a function of $\chi$) will be given by 
\begin{equation}
 \Lambda=-{\Phi\over 2}{dV\over d\Phi}-V\, . \end{equation}
It is the ratio of these two quantities that determines the extrinsic 
``wiggle'' speed $c_{_{\rm E}}$, while the longitudinal (sound type) 
perturbation speed $c_{_{\rm L}}$ will be determined by a 
differential relation: acccording to (\ref{speeds}) we obtain
\begin{equation}
c_{_{\rm E}}^{\,\pm 2}={\Lambda\over {\cal L}}\, ,\hskip 1 cm
c_{_{\rm L}}^{\,\pm 2}= 1+4{w d\Phi\over \Phi dw}\, ,\end{equation}
where the sign is taken to be positive, $\pm=+$ in the spacelike
current regime where $w>0$, and to be negative $\pm=-$ in the timelike
current regime where $w<0$. It is to be observed that in each of these 
regimes the causality restriction $c_{_{\rm L}}^{\,2}\leq 1$ (i.e. the
restriction that the ``sound'' speed should not excede the speed of
light) implies the monotonicity condition $d\Phi/dw\leq 0$, so that
as  corollary it can be seen that we shall always have $0<\Phi\leq
\Phi_{\!_0}$ in the spacelike current regime and $\Phi_{\!_0}\leq \Phi$
in the timelike current regime.

\section{ Noteworthy examples}
\label{examples}

\subsection{Zeroth (chiral) model}

For describing the effect of a current produced by the Witten mechanism in 
the bosonic case, five kinds of simplified but increasingly accurate kinds 
of elastic string model have been developed over the years\cite{CP95}. The 
{\it first} of these is the original Witten model given by (\ref{4}), 
while the {\it second}, {\it third}, {\it fourth}, and {\it fifth} kinds 
can all be expressed in the form (\ref{0}) with simple explicit expressions 
for the function $V$ that will be listed below.

Before proceeding to do so however, it is important to mention the 
degenerate special case of what can be appropriately listed as the
{\it zeroth} model, namely the ``chiral'' case~\cite{CP99} for which the
potential function $V$ is just a constant, 
\begin{equation}  V=m^2\, ,
 \label{0V}\end{equation}
where, as in (\ref{4}), $m$ is a constant  having the dimensions of mass.
(This Kibble type mass parameter $m$ can in practice be expected to be 
of the same order of magnitude as the mass of the Higgs field responsible 
for the vacuum degeneracy with respect to which the cosmic strings under 
consideration arise as vortex defects.) Invariance of the action with 
respect to free variations of the auxiliary field $\Phi$ evidently entails 
that the current in such a model is restricted to satisfy the nullity 
condition 
\begin{equation} 
\varphi_{\vert a}\varphi^{\vert a}=0\, . \end{equation}
It is this ``zeroth'' 
model that is appropriate for describing the special fermionic case 
mentionned above, in which the only occupied states are zero-modes with a 
unique (according to convention exclusively left moving or exclusively 
right moving) orientation. In this effectively self dual case, for which 
it can be seen from (\ref{Lamb}) that -- since the current magnitude 
$\chi$ vanishes -- there is no difference between $\Lambda$  and 
${\cal L}$, the stream function~\cite{C97} $\tilde\varphi$ will be 
identifiable (for suitable normalisation) with the phase potential $\varphi$ 
of which it is the dual. (In much of the relevant literature the stream 
function $\tilde\varphi$ is denoted for simplicity by the letter $\psi$, but
since this usage is redundant in the self dual ``chiral'' case the symbol
$\psi$ is available therein for other purposes, and has been 
used~\cite{CP99} instead for the dilatonic amplitude that is denoted here 
by $\Phi$.)

\subsection{First (Witten) model}
  
It is to be remarked that the Lagrangian (\ref{4}) for the
{\it first} model  is obtainable from that of the {\it zeroth} 
(``chiral'') model with fixed potential (\ref{0V}), 
simply by replacing the variable amplitude $\Phi$ by a constant value, 
i.e. imposing a restraint of the form
\begin{equation}\Phi=\Phi_{\!_0}\, ,\end{equation}
thereby relaxing the nullity restriction. For the non-degenerate Witten model 
obtained in this way ``sound'' travels at the speed of light, i.e
$c_{_{\rm L}}=1$.  (This is the string analogue of the Zeldovich model 
representing the ``stiff'' limit case for an ordinary fluid.)

In so far as the bosonic case is concerned, Peter's analysis~\cite{Peter92a}
has shown, as remarked above,  that the Witten mechanism leads to behaviour
that is not correctly described by this Witten string model (\ref{4}),
since -- unlike what occurs in this {\it first} string model -- it turns out
that $c_{_{\rm L}}$ will really be not just less than $1$ but even less than 
the ``wiggle'' speed $c_{_{\rm E}}$, at least when the current amplitude is 
small.

\subsection{Second (transonic) model}

The simplest case for which both kinds of perturbation are subluminal
is the ``transonic'' case for which they are the same, 
$c_{_{\rm L}}=c_{_{\rm E}}=1$.  A model of this {\it second} type has 
the  convenient property that -- unlike the {\it first} (Witten) kind, but 
like the {\it zeroth} (``chiral'') kind~\cite{CP99}) itsdynamical equations 
are exactly integrable in a flat spacetime background and it can be shown 
using this property ~\cite{C90,C95} or otherwise~\cite{Vilenkin90,Martin95} 
that this `transonic'' model can provide a good description of the averaged 
effect of small wiggles in an underlying string of the simple Nambu Goto 
kind.  In the traditional formulation, the Lagrangian for this  
 ``transonic'' string model is expressible in the form
\begin{equation}
 {\cal L}=-m\sqrt{m^2+w}\, \end{equation}
in which $m$ is a fixed mass parameter as introduced above. In this case 
the transition to the dilatonic formulation is made by taking
\begin{equation}
 \Phi^2=  {m\Phi_{\!_0}^2\over \sqrt {m^2+w}}\, .\end{equation}
The result for the potential function $V$ is in this {\it second}
(``transonic'') case is thereby obtained from (\ref{V}) in the form
\begin{equation}
 V= {m^2\over 2}\Big({\Phi^2\over\Phi_{\!_0}^2}+
 {\Phi_{\!_0}^2\over \Phi^2}  \Big)\, .\label{2V}\end{equation}

\subsection{Third (polynomial) model}
                                            
For the purpose~\cite{CP95} of representing the effect of a current 
arising from Witten's ``superconductivity'' mechanism~\cite{Witten85}, a
more accurate description, at least in the weak current limit, can
be obtained by using a {\it third} kind of model having a Lagrangian
function of the polynomial form
\begin{equation} 
{\cal L} =-m^2-{w\over 2}\Big( 1+{w\over m_\star^{\,2}}\big)
\, ,\label{third}\end{equation}
in which as well as the original ``Kibble'' mass parameter $m$ an
independent ``Witten'' constant mass parameter $m_\star $ is also involved.
In a model of this {\it third} kind the transition to the dilatonic
formulation is made by taking
\begin{equation} \Phi^2=  \Phi_{\!_0}^2\Big(1-{2w\over m_\star^{\,2}}\Big)\,
 ,\end{equation}
and the corresponding formula for the potential function is
\begin{equation}
 V = m^2+{m_\star^{\,2}\over 8}\Big({\Phi^2\over \Phi_{\!_0}^2}
-1\Big)^2\, .\label{3V}\end{equation}
  
\subsection{Fourth (rational) model}

On the assumption that the $m_\star $ is relatively small compared with
$m$ the preceding model will indeed be characterised by supersonic wiggle
propagation, $c_{_{\rm E}}>c_{_{\rm L}}$, in the weak current 
($w\rightarrow 0$) limit, in accordance with results of detailed 
numerical, analysis~\cite{Peter92a} of the Witten mechanism, but to 
obtain a model that gives a qualitatively realistic description for 
larger currents, at least in the spacelike case, $w>0$, a formula of 
the non polynomial but still rational form
\begin{equation}
{\cal L}=-m^2-{w\over 2}\Big(1+{w\over m_\star^{\,2}}\Big)^{-1} 
\ ,\label{fourth}\end{equation}
was found to be more satisfactory. In a model of this {\it fourth}
kind the transition to the dilatonic formulation is made by taking
\begin{equation}
 \Phi^2=  \Phi_{\!_0}^2\Big(1+{w\over m_\star^{\,2}}\Big)^{-2}\, ,\end{equation}
and the corresponding formula for the potential function is
\begin{equation}
 V = m^2+{m_\star^{\,2}\over 2}\Big({\Phi\over\Phi_{\!_0}}-1 \Big)^2
\, .\label{4V}\end{equation}
It is to be remarked that although the Lagrangian (\ref{fourth}) is
more complicated than its polynomial predecessor (\ref{third}) the
ensuing merely quadratic formula (\ref{4V}) for the potential is 
actually simpler than its quartic predecessor (\ref{3V}).

\subsection{Fifth (logarithmic) model}

To obtain a qualitatively satisfactory description not only in the
spacelike but also the timelike current regime (where $w<0$)
it was found~\cite{CP95} to be necessary resort to the use of
a non-rational equation of state of the form
\begin{equation}
 {\cal L}= -m^2-m_\ast^{\,2}\, {\rm ln}\Big\{1+{w\over m_\ast^{\,2}}
\Big\}\, \end{equation}
where $m_\ast$ is another fixed mass parameter. (In order for this to 
agree in the weak field limit with the preceeding third and fourth kinds
of model, this new mass parameter would have to be related to that of 
these preceding examples by  $2m_\ast^{\,2}=m_\star^{\,2}$.) In a model 
of this {\it fifth} -- and, for the treatment of the Witten mechanism, 
most realistic -- kind, the transition to the dilatonic formulation is 
made by taking
\begin{equation}
 \Phi^2=  \Phi_{\!_0}^2\Big(1+{w\over m_\ast^{\,2}}\Big)^{-1}\,
 ,\end{equation}
and the corresponding formula for the potential function is
\begin{equation}
 V = m^2+{m_\ast^{\,2}\over 2}\left({\Phi^2\over \Phi_{\!_0}^2}-1
- {\rm ln}\Big\{ {\Phi^2\over \Phi_{\!_0}^2}\Big\} \right)\, . 
\end{equation}  

%

\end{document}